\RequirePackage[T1]{fontenc}
\documentclass[11pt,a4paper,titlepage]{article}

\usepackage[dvipsnames,table]{xcolor}	

\usepackage{lmodern} 
\usepackage[mono=false]{libertine}

\usepackage{geometry}
\usepackage[main=english,ngerman]{babel}
\usepackage{hyperref}
\usepackage{sectsty}
\usepackage{graphicx}
\usepackage{enumitem}
\usepackage{amsmath}
\usepackage{wrapfig}
\usepackage{multirow}

\usepackage{float}
\usepackage{array,booktabs}
\usepackage{tabularx}
\usepackage{hhline}

\usepackage[
	skip	= 0.5em,
	indent	= 0pt
]{parskip}

\definecolor{PrimaryBlue}{HTML}{3C6382}
\sectionfont{\color{PrimaryBlue}}
\subsectionfont{\color{PrimaryBlue}}
\subsubsectionfont{\color{PrimaryBlue}}
\paragraphfont{\color{PrimaryBlue}}

\definecolor{MarkGreen}{HTML}{58bf36}
\definecolor{BGTable}{HTML}{f4f5f7}

\definecolor{WPOneColor}{HTML}{e3fcef}
\definecolor{WPTwoColor}{HTML}{deebff}
\definecolor{WPThreeColor}{HTML}{ffebe6}

\usepackage{pifont}
\def\check{\textcolor{MarkGreen}{\ding{55}}}
\newcolumntype{?}{!{\vrule width 1pt}}

\newcolumntype{a}{>{\columncolor{WPOneColor}}c}
\newcolumntype{b}{>{\columncolor{WPTwoColor}}c}
\newcolumntype{d}{>{\columncolor{WPThreeColor}}c}

\makeatletter
\def\hlinewd#1{%
\noalign{\ifnum0=`}\fi\hrule \@height #1 %
\futurelet\reserved@a\@xhline} 
\makeatother

\usepackage[
	backend		= biber,
	style		= numeric,
	giveninits	= true,
	defernumbers= true, 
	maxbibnames	= 6,
	minbibnames	= 3,
	hyperref    = true,
    sorting     = anyt
]{biblatex}

\makeatletter
\let\blx@rerun@biber\relax
\makeatother

\makeatletter
\newcommand\Setmaxbibnames[1]{\renewcommand\blx@maxbibnames{#1}}
\newcommand\Setminbibnames[1]{\renewcommand\blx@minbibnames{#1}}
\makeatletter

\usepackage{dfgreport}

\title{%
	Methods and Tools to Advance the Retrieval of Mathematical Knowledge from Digital Libraries for Search-, Recommendation- and Assistance-Systems
}
\author{Bela Gipp, Andr\'{e} Greiner-Petter, Moritz Schubotz, Norman Meuschke}

\reportdate{March 2023}
\dfgref{GI 1259-1} 
\projectid{350192710}
\projecttitle{%
	Methoden und Werkzeuge zur Verbesserung des Zugriffs auf\newline 
	mathematisches Wissen in Digitalen Bibliotheken für\newline
	Such-, Empfehlungs- und Assistenzsysteme%
}
\applicant{Prof.~Dr.~Bela Gipp}
\applicantaffil{%
	University of Göttingen\newline
        Dept. of Computer Science\newline
	Scientific Information Analytics Group\newline 
	Papendiek 14, 37073 Göttingen%
}
\fundingperiod{July 1\textsuperscript{st}, 2018 -- December 31\textsuperscript{st}, 2022}

\nocite{*}

\begin{document}
\pagenumbering{roman}
\makeatletter
\begin{titlepage}
\parindent 0 pt
\newgeometry{top=4.2cm,bottom=4.2cm,right=0.8in,left=0.8in}
    \begin{center}
       {\Large \textcolor{PrimaryBlue}{\textbf{Final Report for the DFG-Project \glqq MathIR\grqq}} \par}
       {\rule{2cm}{0.7pt}}
       
       {\Large \textcolor{PrimaryBlue}{\textbf \@title} \par}
       \vspace{2em}
       {\large \@author \par}
       \vspace{2em}
       {\@reportdate}
   \end{center}
   \vfill
   
   {\textcolor{PrimaryBlue}{\textbf{Reference Number:}}} {\@dfgref\par}
   {\textcolor{PrimaryBlue}{\textbf{Project Number:}}} {\@projectid\par}
   \vspace{1em}
   
   {\textcolor{PrimaryBlue}{\textbf{Applicant}}\par}
   {\@applicant \newline}
   {\@applicantaffil \par}
   \vspace{1em}
   
   {\textcolor{PrimaryBlue}{\textbf{German Project Title}}\par}
   {\@projecttitle \par}
   \vspace{1em}
   
   {\textcolor{PrimaryBlue}{\textbf{Reporting Period}}\par}
   {\@fundingperiod \par}
   \vspace{1em}

\end{titlepage}
\makeatother

\section*{Abstract}\label{sec:conclusion} 
This project investigated new approaches and technologies to enhance the accessibility of mathematical content and its semantic information for a broad range of information retrieval applications. 
To achieve this goal, the project addressed three main research challenges: (1) syntactic analysis of mathematical expressions, (2) semantic enrichment of mathematical expressions, and (3) evaluation using quality metrics and demonstrators. To make our research useful for the research community, we published tools that enable researchers to process mathematical expressions more effectively and efficiently.


The project has made significant research contributions to various Mathematical Information Retrieval (MathIR) tasks and systems, including plagiarism detection  and recommendation systems, search engines, the first mathematical type assistance system, math question answering and tutoring systems, automatic plausibility checks for mathematical expressions on Wikipedia, automatic computability of mathematical content via Computer Algebra Systems (CAS), and others. Although our project focused on MathIR tasks, its impact on other natural language research was significant, leading to a more extensive range of demonstrators than originally expected. Many of these demonstrators introduced novel applications, such as the tutoring system \href{https://physwikiquiz.wmflabs.org/}{PhysWikiQuiz}~~\cite{ScharpfSSG22} or \href{https://lacast.wmflabs.org/}{LaCASt}~\cite{GreinerPetterCYS22}, which automatically verifies the correctness of math formulae on Wikipedia or the Digital Library of Mathematical Functions (DLMF) via commercial CAS.


During the project, we published 29 peer-reviewed articles in international venues, including prestigious conferences like the \href{https://jcdl.org/}{\textit{Joint Conference on Digital Libraries (JCDL)}}~\cite{MeuschkeSSK19a,ScharpfSG22,ScharpfSYH20a,SchubotzGMT20a,SchubotzGSM18a,SchubotzSDN18} and \href{https://thewebconf.org/}{\textit{The Web Conference (WWW)}}~\cite{GreinerPetterSMB20a,ScharpfSG21b} (\href{http://portal.core.edu.au/conf-ranks/}{CORE rank A*}), as well as journals such as \href{https://www.computer.org/csdl/journal/tp}{\textit{IEEE Transactions on Pattern Analysis and Machine Intelligence (TPAMI)}}~\cite{GreinerPetterSBS22} (IF: 24.314) and \href{https://www.springer.com/journal/11192}{\textit{Scientometrics}}~\cite{GreinerPetterYRM20a} (IF: 3.801). Our Wikipedia demonstrator was also featured in public media. Furthermore, we actively presented our contributions, especially demonstrators, to the research community in multiple workshops~\cite{GreinerPetterRSA19a,PetersenSG18a,ScharpfSCG19a,ScharpfSG18a,ScharpfSG21a,SchubotzSDN18,AsakuraGAM20,ScharpfSSG22}.


This project has strengthened our international collaborations, particularly with colleagues at the National Institute of Standards and Technology (NIST) in the US and the National Institute of Informatics (NII) in Japan. Several subprojects were partially developed in course projects and theses at the Universities of Konstanz, Wuppertal, and Göttingen, exposing junior researchers to cutting-edge technologies and sensitizing students and researchers to the outstanding issues in MathIR technologies. We firmly believe that this project will have a lasting effect on following MathIR technologies. Several of the subprojects initiated as part of this grant are ongoing and motivating follow-up DFG projects, such as \href{https://gepris.dfg.de/gepris/projekt/437179652?language=en}{\textit{Analyzing Mathematics to Detect Disguised Academic Plagiarism} (project no. 437179652)}.

\pagebreak
\section*{Most Influential Project Publications}
\subsection*{Publications with Scientific Quality Assurance}
\newrefcontext[sorting=ydnt]
\printbibliography[keyword=top,heading=none]\label{sec:pubs-top}
\begin{center}
A complete list of publications resulting from this DFG project is available at\\
\href{https://gipp.com/pub/#mathir}{\texttt{gipp.com/pub/\#mathir}}.
\end{center}

\pagebreak
\tableofcontents

\pagebreak
\pagenumbering{arabic}
\section{Progress Report}\label{sec:report}


\subsection{Background and Project Objectives}
Information Retrieval (IR) systems have become indispensable for finding relevant information despite today's ubiquitous information overload. Web search engines like Google and Bing, and the recommender systems embedded within Amazon and Netflix are prominent examples of the many domain-specific systems for indexing and accessing content we use daily. In academia, finding relevant literature is a vital task in all research disciplines but the exponential growth in the number of publications makes it increasingly challenging. Consequently, IR systems also play a crucial role in facilitating access to scholarly literature.

Despite IR systems' importance in accessing scholarly literature and their dependence on accessible information, much essential data remains unused, particularly non-textual data, such as images, audio and video data, and mathematical content. Recent research has focused on improving the accessibility of information in non-textual data, but mathematical content has been largely neglected. However, researchers, especially in STEM\footnote{Science, Technology, Engineering, and Mathematics.} fields, often communicate critical information via mathematical expressions. At best, ignoring mathematical content can be confusing. More likely, it renders the content useless.

Existing IR systems for mathematical content rely on visual resemblance and rarely access semantic information on mathematical expressions. Therefore, the goal of this DFG project was to research new approaches and technologies to automatically make semantic information in mathematical expressions accessible for a wide range of IR applications.

To achieve this goal, we addressed three major issues: (1) the syntactic analysis of mathematical expressions, (2) the semantic enrichment of mathematical expressions, and (3) the evaluation using quality metrics and demonstrators. To make our research accessible and useful for the research community, we published tools that enable researchers to process mathematical expressions more effectively and efficiently. This approach was inspired by well-established tools for natural language processing tasks, such as Stanford's Natural Language Processing Toolkit. We aimed to provide similarly flexible processing engines for mathematical expressions.

\subsection{Project Results}\label{sec:results}

Hereafter, we explain the projects we completed as part of the grant. As most of these projects have contributed towards multiple research objectives, we have organized the report into project-centered descriptions to provide a better understanding of the overall progress and results. Each section of the report briefly summarizes how the project has contributed towards the grant's overall goal, followed by a summary of the project and its results. For a more detailed overview of the project's contribution towards the research objectives, please refer to Appendix~\ref{sec:contribution-table}. Most of the projects have implemented their own demonstrators. The following list shows all the demonstrators that have been developed as part of this grant:

\begin{itemize}
\item PhysWikiQuiz~\cite{ScharpfSSG22}: A question-answering system for physics-related questions.
\item MathQA~\cite{ScharpfSG22,SchubotzSDN18}: A question-answering system for math-related questions (laid the foundation for PhysWikiQuiz).
\item AnnoMathTeX~\cite{ScharpfSG21b,ScharpfMSB19a}: A tool to annotate mathematical LaTeX expressions with mathematical concepts.
\item MathMLben~\cite{SchubotzGSM18a}: A benchmark for generated MathML data.
\item MathMLTools~\cite{GreinerPetterSCG18a}: A development toolset that improves handling MathML data in Java and provides numerous interfaces for typical tasks, such as conversions, similarity calculations, and data compliance tests.
\item LaCASt~\cite{CohlGS18a,GreinerPetterSCG19a,GreinerPetterCYS22,GreinerPetterSBS22}: A conversion tool that translates mathematical LaTeX into the syntax of Computer Algebra Systems.
\item Mediawiki Extensions~\cite{SchubotzGMT20a}: Tools that enhance the semantics of mathematical expressions in Mediawiki applications, such as Wikipedia.
\item DLMF/DRMF~\cite{GreinerPetterCYS22}: The translations of the LaCASt project shall be added to the DLMF/DRMF; the associated evaluations helped to detect mathematical errors in the DLMF.
\end{itemize}

\subsubsection{Fundamental MathIR Contributions}

This section focuses on projects that addressed fundamental problems in MathIR systems. An issue that most mathematical data handling systems exhibit is that they either consider individual symbols or the entire expression but neglect significant and meaningful subexpressions. Compound mathematical expressions contain vital semantic information that is lost when logical components, such as function calls, arguments, parameter structures, and arithmetic logic, are ignored. The composition of mathematical formulae has received little attention from the NLP community, mainly because identifying meaningful subexpressions is context-dependent. 

Another fundamental issue is the lack of a ground truth~\cite{AsakuraGAM20} for components of mathematical expressions which hampers the development of machine learning applications in the MathIR community. A major issue for creating such a large annotated dataset is the lack of a unified standard and the open question if such a ground truth can actually exist. 
Our work addressed both fundamental issues by studying mathematical objects of interest and developing the AnnoMathTex and MathMLben tools to further develop standardized annotations.

\paragraph{Mathematical Objects of Interest (MOI)} 

Mathematical Information Retrieval (MathIR) systems face a common issue---they cannot differentiate between important mathematical expressions, such as functions, and less important or replaceable information, such as variables. In contrast, natural language processing approaches use stop word removal to focus on semantically essential text parts. An equivalent method does not exist for mathematical expressions. To discover the mathematical objects of interest (MOI)~\cite{GreinerPetterSMB20a}, we analyzed mathematical notations in the two largest scientific datasets of mathematical documents: arXiv and zbMATH Open. Our analysis of 2.5 billion mathematical expressions revealed that mathematical notations follow a frequency distribution pattern similar to that of words in natural languages.

The discovery of this pattern allowed us to use similar techniques as in natural languages to measure the significance of mathematical expressions and remove potential mathematical \textit{stop words}. One of the most well-known schemes to measure significance based on frequency distributions is the term frequency-inverse document frequency measure (TF-IDF). By applying identical calculations to mathematical expressions, we created a novel context-sensitive math search system, enabled the first assistance system for mathematical inputs, and provided useful information for plagiarism detection and scientific recommender systems.

\paragraph{Grounding of Formulae}
\begin{figure}
    \centering
    \includegraphics[width=\textwidth]{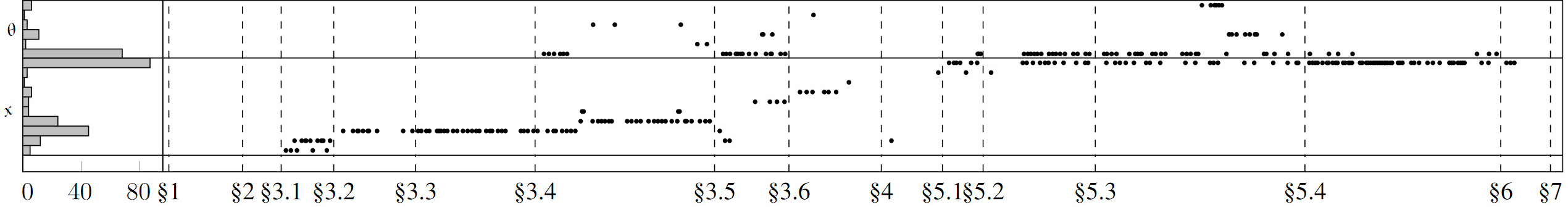}
    \caption{Different semantic annotations of $\theta$ and $x$ within the same article.}
    \label{fig:x-dist}
\end{figure}

The MathIR community often neglects that the semantics of mathematical notations can change rather frequently, even within the same document. While authors can redefine variables, constants, and functions at their discretion (e.g., $\pi$), little is known about the extent of this phenomenon. Mathematical language follows certain notation styles and rules that have developed over thousands of years. We expected that mixing semantics for identical expressions would be rare, at least within qualitative research articles. 

To test this assumption and ground formulae in mathematical objects, we manually annotated mathematical objects with mathematical concepts derived from the context of the objects~\cite{AsakuraGAM20}. Figure~\ref{fig:x-dist} illustrates how the mathematical concepts for $\theta$ and $x$ change multiple times within one article, and sometimes even change back and forth within a single subsection. Although one might expect such erratic semantic changes for heavily overloaded identifiers like $x$, we also observed this behavior for rather specific, less frequent identifiers like $\theta$.

For instance, the connection between Euler numbers and Euler polynomials demonstrates how $E_n$ can change its semantic meaning within the scope of a single formula:
\begin{equation}
E_n = 2^n E_n\left( \tfrac{1}{2} \right)
\end{equation}

Our study~\cite{AsakuraGAM20} showed that the scope of mathematical semantic information can be very narrow and change back and forth within subsections, paragraphs, or other context windows.

\paragraph{AnnoMathTex}

The lack of high-quality annotated datasets for mathematical literature is a major challenge for MathIR. To address this issue, we developed AnnoMathTex~\cite{ScharpfSG21b,ScharpfMSB19a}, a system that provides artificial intelligence (AI) guidance to improve the workflow of curating annotated datasets for mathematical literature. AnnoMathTex can recommend annotations for Wikipedia articles, drawing from arXiv, Wikipedia, Wikidata, the text surrounding the formula to be annotated, and previous user-made annotations. The system further expands these sources through fuzzy searches and linked Wikidata properties. To address the unknown scope for annotation, AnnoMathTex distinguishes global from local annotations. Our evaluation of AnnoMathTex demonstrated that the system accelerates the process of manually annotating a dataset by a factor of 1.4 for entire formulae and 2.4 for single identifiers. Moreover, we employed the system to automatically annotate Wikipedia articles and update properties in Wikidata. Remarkably, 80\% of these changes were accepted in Wikipedia and 67\% in Wikidata. Overall, AnnoMathTex shows great potential to significantly advance the field of MathIR by accelerating the process of annotating and updating Wikidata properties.

\paragraph{MathMLben}

The MathMLben project~\cite{SchubotzGSM18a} addressed the lack of standard datasets for well-formatted and semantically enhanced content MathML. This benchmark contained 305 formulae taken from English Wikipedia articles, which we semantically enriched manually to obtain error-free and accurate MathML data. As a follow-up, we evaluated state-of-the-art LaTeX to MathML conversion tools to verify their accuracy and practicality in generating MathML. This evaluation was a crucial step forward for all MathIR-related tasks because MathML is hardly ever written manually, due to its convoluted XML structure. Typically, conversion tools are employed to generate MathML from LaTeX or image sources. However, we found that none of the tools achieved sufficient accuracy, and only three were able to generate content MathML, while all others exclusively provided presentation MathML with varying accuracy.


Our analysis identified the lack of context awareness as the main weakness of all conversion tools. Mathematical expressions are highly context-dependent and require substantial fundamental knowledge, often referred to as common knowledge. For example, the simple expression $\pi(x+y)$ may represent a multiplication between the mathematical constant $\pi$ and $x+y$. However, the formula could also occur in a number theory context and discuss the number of primes, in which case $\pi$ more likely refers to the prime counting function $\pi(n)$. Without context analysis, no tool could distinguish one from the other, although the difference is crucial.

The MathMLben dataset was designed to be extendable and has been used for follow-up projects. Most noteworthy are the extensions for Wikimedia~\cite{ScharpfSCG19a,ScharpfSG21a,ScharpfSG21b,ScharpfSSG22}.

\subsubsection{LaCASt}

The LaTeX to Computer Algebra Systems translator (LaCASt)~\cite{CohlGS18a,GreinerPetterSCG19a,GreinerPetterCYS22,GreinerPetterSBS22} was a major project of this DFG grant. LaCASt is the first context-sensitive translator that analyzes the structure of mathematical LaTeX inputs and considers the textual context of a formula to disambiguate mathematical expressions. This translation process requires solutions to all objectives of the DFG grant, i.e., a syntactic analysis of mathematical expressions, a semantic enrichment pipeline, and novel quality metrics and evaluation techniques.


To achieve this, LaCASt first builds a dependency graph for mathematical notations within a document to link relevant document sections and retrieve semantic information about specific formulae. For example, let us assume the formula $\pi(x+y)$ has previously been introduced in the document as the prime counting function. The created link in the graph allows LaCASt to retrieve the relevant information and disambiguate $\pi$. The next step is to annotate nodes in the graph with textual descriptions surrounding the formula. These textual annotations are used to retrieve standard notations from the Digital Library of Mathematical Functions (DLMF). Many well-known formulae, such as the prime counting function $\pi(n)$, have standardized notations. The standard notation is essential to link the mathematical concept, here 'prime counting function', to the relevant subexpression, such as $\pi(\cdot)$. The standard notations from the DLMF helped to identify the logical syntax of an expression and annotate the syntactic elements with semantic information retrieved from the context.


LaCASt then tries to replace the general LaTeX expressions, such as \verb|\pi(n)|, with semantically enhanced LaTeX taken from the DLMF, here \verb|\nprimes@{n}| by considering the gathered information. After the disambiguation, the expression can be mapped to the syntax of a Computer Algebra System (CAS). For example, a translation to Mathematica would map \verb|\nprimes@{n}| to \verb|PrimePi[n]|. In our study, the only option to determine if a translation was correct, was to consult an expert for both the mathematical formulae and the CAS. To evaluate LaCASt, we created a dataset of 95 formulae, which we randomly selected from Wikipedia articles and manually translated to two CAS syntaxes (Maple and Mathematica) with the help of an expert. LaCASt correctly translated 27\% of the formulae. Notably, the expert could only translate 81\% of the formulae, which underscores the task's complexity and indicates a possible upper bound. A more comprehensive common knowledge dataset could have improved LaCASt's performance by 20\%. Nonetheless, LaCASt outperformed the state-of-the-art baseline---Mathematica's LaTeX import function---which only translated 9\% of the formulae correctly.

Another option to check the correctness of translations, which we evaluated in our study, is to compute the translated formula in the target Computer Algebra System and analyze the results. The equations should originate from a reliable source that established the equation's correctness---we used the DLMF. If the translated equation is invalid, there are three possible explanations: (1) the source equation was incorrect, (2) the translation was incorrect, or (3) the CAS exhibits a bug. Given that the source of the equation and the target CAS (in our study Mathematica or Maple) are highly reliable, reasons (1) and (3) are unlikely.

LaCASt could translate 72\% of the DLMF to the two CAS and evaluated 48\% of the 4,713 translated equations successfully. The other 52\% of the equations could not be evaluated due to missing semantic information, such as branch cut positions, domains, and other constraints. Thus, a failed verification still required a manual investigation. However, we can conclude that the translations performed by LaCASt are reliable. Interestingly, the approach could detect errors in the DLMF and the two major CAS, Maple and Mathematica.


Recently, we demonstrated the potential of LaCASt by verifying Wikipedia edits. Although LaCASt is not yet productive, a demo page showcasing LaCASt's capabilities is available at \href{https://tpami.wmflabs.org}{\texttt{tpami.wmflabs.org}}.

\subsubsection{DLMF \& DRMF}

The LaCASt project also contributed to improving the Digital Library of Mathematical Functions and the Digital Repository of Mathematical Formulae (DRMF). The evaluation approach of computing translated formulae in the target CAS (see above) allowed the detection of erroneous mathematical content within the prestigious DLMF. The various identified errors included missing semantic information, incorrect links, sign errors, and others.

The DLMF is manually written using semantic LaTeX macros, which enables LaCASt to achieve optimal performance as the typically uncertain disambiguation of formulae can be skipped in most cases. This benefit of the DLMF allows for reliable translations of formulae to widely used CAS such as Mathematica and Maple, which can be very beneficial for other researchers.

We are currently discussing the integration of LaCASt into the DLMF or the expansion of the DRMF with the National Institute of Standards and Technology (NIST). A demo page showcasing all translations and evaluation results for the DLMF is available at \href{https://lacast.wmflabs.org}{\texttt{lacast.wmflabs.org}}.

\subsubsection{Wikipedia and Wikidata Extensions \& Projects}

\begin{wrapfigure}{l}{.5\textwidth}
	\centering
    \vspace{-0.2cm}
	\includegraphics[width=0.5\textwidth, trim=0cm 0cm 0cm 0cm, clip]{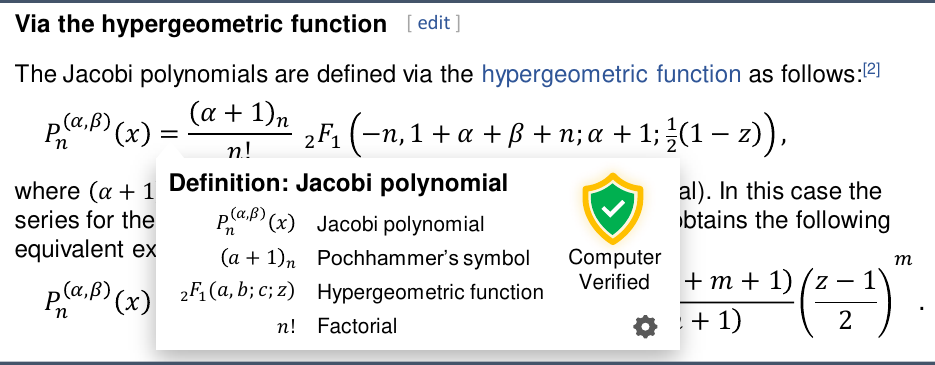}
    \vspace*{-0.7cm}
	\caption{Mathematical semantic annotation in Wikipedia.}
	\label{fig:wikipedia-popup}
    \vspace{-0.1cm}
\end{wrapfigure}

We focused on creating demonstrators primarily for two systems within the Mediawiki platform---the encyclopedia Wikipedia and the knowledgebase Wikidata. For Wikipedia, we realized several extensions that enable providing semantically enhanced mathematical content as part of the encyclopedia~\cite{SchubotzGMT20a}. One of the extensions enables users to annotate mathematical formulae in Wikipedia articles with items from Wikidata to provide additional semantic information. For example, the famous formula $E=mc^2$ in the English Wikipedia article on mass-energy equivalence is now linked with the Wikidata item \verb|Q35875|. This linking enables providing additional information on the formula to the end-user directly within the Wikipedia article. When someone hovers over an annotated mathematical formula, a popup appears with information such as the name, a description, and a list of elements and their meaning. In the future, we aim to extend these popups to display verified equations, e.g., checked by LaCASt, as shown in Figure~\ref{fig:wikipedia-popup}.

We also explored potential applications that can be derived from well-maintained knowledge-base systems such as Wikidata~\cite{ScharpfSG21b}. We elaborated on the possibility of using Wikidata items to automatically generate OpenMath content dictionaries, which are required to annotate more complex mathematical concepts in content MathML data~\cite{Schubotz18a}. The ability to link mathematical expressions and concepts with natural language expressions (which is possible in the case of Wikidata, even for multi-lingual expressions) is beneficial for a large variety of MathIR systems. 

Hereafter we explain two projects that emerged from this idea---PhysWikiQuiz and MathQA.

\subsubsection{PhysWikiQuiz \& MathQA}

PhysWikiQuiz~\cite{ScharpfSSG22} and its predecessor MathQA~\cite{ScharpfSG22,SchubotzSDN18} are systems that use Wikidata to automatically generate mathematical questions. Wikidata is a knowledge graph with items connected via a large variety of properties. It can be exploited to generate math-related questions, e.g., for question-answering systems and educational purposes. The PhysWikiQuiz system first takes an example formula from a Wikidata item, such as $v = s/t$ (speed). It then retrieves the names and units of each symbol involved in the formula via Wikidata properties. Using a computer algebra system (in this case, SymPy), the system rearranges the formula and generates test calculations that can be presented to the user and verified by the computer algebra system. For instance, given the input speed', the system may generate the question: \texttt{What is the distance $s$, given speed $v = 10ms^{-1}$ and duration $t=6s$?}, with an expected correct answer of $60m$. The system also checks the value and unit of the answer.

\subsection{Project Adjustments}\label{sec:adjustments}

Significant adjustments to the research agenda were not necessary but the project team had to relocate multiple times, moving from the University of Konstanz to the University of Wuppertal, and later to the University of Göttingen. These relocations caused staff shortages, which required applying for an extension of the originally planned project timeline. While the extension was necessary, it also provided opportunities that would have been difficult to achieve otherwise, such as collaborating with new partners from new institutions.

\subsection{Commercial Viability}\label{sec:commercial}


Although some of our projects, such as PhysWikiQuiz and LaCASt, have the potential for commercial exploitation, we do not plan to use the project results commercially. We are committed to making the findings and systems resulting from our research accessible to everyone free of charge as we firmly believe that open access is crucial for advancing research and innovation.

\subsection{Follow-up Projects}

The improved accessibility and retrievability of mathematical knowledge achieved in this project have enabled research on downstream applications. 

One such application is \textit{Math-based plagiarism detection (MathPD)}, i.e., the search for mathematical content in academic documents that is similar to the content in prior publications without that being justified and acknowledged according to academic standards. MathPD poses numerous application-specific questions and challenges that we investigate in the DFG project \href{https://gepris.dfg.de/gepris/projekt/437179652?language=en}{\textit{Analyzing Mathematics to Detect Disguised Academic Plagiarism} (project no. 437179652)}

In addition to enabling checks for plagiarized mathematical content, we seek to improve the plagiarism detection process by eliminating data privacy concerns and the dependence on near-monopolistic service providers. To achieve this goal, we are researching privacy-preserving methods to identify similar content in academic publications, which includes text, images, citations, and mathematical content. Moreover, we investigate the use of distributed ledger technology to create a distributed, trustless system for running privacy-preserving plagiarism checks. We will submit a funding proposal for this research to the DFG in July 2023. 

The technology developed in this project serves as the foundation for the portal of the \href{https://gepris.dfg.de/gepris/projekt/460135501?language=en}{National Research Data Initiative (NFDI) project for mathematics (MaRDI) (project no. 460135501)}. Specifically, the semantically enhanced version of the DLMF formulae was the initial seed of mathematical research data imported to the portal \href{https://portal.mardi4nfdi.de}{\texttt{portal.mardi4nfdi.de}}. In the MaRDI project, we continue our efforts to enhance the semantics of mathematical research data and improve its accessibility and retrievability.

\subsection{Contributors}
This section provides an overview of the primary contributors and project partners.

\subsubsection{Project Employees}
This project supported four research associates: Corinna Breitinger, Dr. Andr\'{e} Greiner-Petter, Dr. Norman Meuschke, and Dr. Moritz Schubotz.

\subsubsection{Partner Institutions}
Besides the involved host institutions, the universities of Konstanz, Wuppertal and G\"{o}ttingen in Germany, we collaborated with four international institutions:
\begin{enumerate}
    \item FIZ Karlsruhe - Leibniz Institute for Information Infrastructure (zbMATH Open), Berlin, Germany
    \item National Institute of Informatics (NII), Tokyo, Japan
    \item National Institute for Standards and Technology (NIST), Gaithersburg, USA
    \item Wikimedia Foundation
\end{enumerate}

\subsubsection{Qualification of Junior Researchers}\label{sec:theses}

The following doctoral dissertations have been completed during the project:
\begin{enumerate}
    \item Andr\'{e} Greiner-Petter, Making Presentation Math Computable - A Context-Sensitive Approach for Translating LaTeX to Computer Algebra Systems. Dissertation, University of Wuppertal, 2022.~\cite{GreinerPetter2022b}
    \item Norman Meuschke, Analyzing Non-Textual Content Elements to Detect Academic Plagiarism. Dissertation, University of Konstanz, 2021.~\cite{Meuschke2021}
\end{enumerate}
The following doctoral dissertations are still in progress: 
\begin{enumerate}
    \item Phillipp Scharpf, Mathematical Entity Linking, University of Konstanz
    \item Corinna Breitinger, Academic Recommender Systems, University of Konstanz
\end{enumerate}

In addition, 7 master's and 17 bachelor's theses were completed in relation to the project.





\pagebreak
\subsection{Contributions of Project Publications to Grant Objectives}\label{sec:contribution-table}
Table~\ref{tab:work-packages} shows the project's work packages as defined in the proposal and Table~\ref{tab:contributions} how the publications resulting from this project contributed to the work packages. The proposal (in German) is available at \url{https://doi.org/10.5281/zenodo.7908591} and our project page at \url{https://gipplab.org/projects/mathir/}. 

\begin{table}[H]
\centering
\caption{Short description of work packages.}
\label{tab:work-packages}
\begin{tabularx}{\textwidth}{?l?X?}
    \hlinewd{1pt}
    \rowcolor{BGTable} \multicolumn{1}{?c?}{\textbf{WP}} & \multicolumn{1}{c?}{\textbf{Work Package Description}} \\
    \hlinewd{1pt}

    \rowcolor{BGTable}
    \textbf{1}   & \textbf{Syntactic Analysis of Mathematical Expressions} \\ \hline
    1.1 & Differentiating Math and Non-Math Content \\ \hline
    1.2 & Classification of Mathematical Expressions \\ \hline
    1.3 & Tokenization of Complex Mathematical Expressions \\ \hline
    1.4 & Pattern Matching and Interaction \\ 
    \hlinewd{1pt}

    \rowcolor{BGTable}
    \textbf{2}   & \textbf{Semantic Enrichment with Math Concepts} \\ \hline 
    2.1 & Annotating Math Tokens with Natural Language Tokens \\ \hline
    2.2 & Annotating Math Tokens with Mathematical Concepts \\ \hline
    2.3 & Consistency and Quality Checks \\
    \hlinewd{1pt}

    \rowcolor{BGTable}
    \textbf{3}   & \textbf{Quality Metrics, Demonstrators, and Evaluation} \\ \hline 
    3.1 & Development of Quality Metrics for Mathematical Markups \\ \hline
    3.2 & Demonstrators \\ \hline
    3.3 & Evaluations \\
    \hlinewd{1pt}
\end{tabularx}
\end{table}

\begin{table}[H]
\centering
\setlength\tabcolsep{2.5pt}
\caption{Overview of project publications and their contributions to the work packages.}
\label{tab:contributions}
\footnotesize
\begin{tabularx}{\textwidth}{?c?X?a|a|a|a?b|b|b?d|d|d?}
\hlinewd{1pt}

\rowcolor{BGTable} & \multicolumn{1}{c?}{} & \multicolumn{4}{c?}{\textbf{WP 1}} & \multicolumn{3}{c?}{\textbf{WP 2}} & \multicolumn{3}{c?}{\textbf{WP 3}} \\ \hhline{>{{\vrule width 0.6pt}}|>{\arrayrulecolor{BGTable}}->{{\vrule width 0.6pt}\arrayrulecolor{black}}|>{\arrayrulecolor{BGTable}}->{{\vrule width 0.6pt}\arrayrulecolor{black}}|----|---|---|} 
\rowcolor{BGTable}\multirow{-2}{*}{\textbf{Year}} & \multicolumn{1}{c?}{\multirow{-2}{*}{\textbf{Publication}}} & \multicolumn{1}{c|}{\textbf{1.1}} & \multicolumn{1}{c|}{\textbf{1.2}} & \multicolumn{1}{c|}{\textbf{1.3}} & \textbf{1.4} & \multicolumn{1}{c|}{\textbf{2.1}} & \multicolumn{1}{c|}{\textbf{2.2}} & \textbf{2.3} & \multicolumn{1}{c|}{\textbf{3.1}} & \multicolumn{1}{c|}{\textbf{3.2}} & \textbf{3.3} \\ 
\hlinewd{1pt}

\cellcolor{BGTable} & \fullcite{GreinerPetterSBS22}; \cite{GreinerPetterSBS22} & 
    \check & \check & \check & \check & 
        \check & \check & \check & 
            \check & \check & \check \\ \hhline{>{{\vrule width 0.6pt}}|>{\arrayrulecolor{BGTable}}->{{\vrule width 0.6pt}\arrayrulecolor{black}}|-----------}
\cellcolor{BGTable} & \fullcite{GreinerPetterCYS22}; \cite{GreinerPetterCYS22} &  & \check & \check &  & \check & \check &  & \check &  & \check \\ \hhline{>{{\vrule width 0.6pt}}|>{\arrayrulecolor{BGTable}}->{{\vrule width 0.6pt}\arrayrulecolor{black}}|-----------}
\multirow{-3}{*}[5em]{\cellcolor{BGTable}\textbf{2022}} & \fullcite{ScharpfSG22}; \cite{ScharpfSG22} &  &  &  &  & \check & \check & \check &  &  &  \\ 
\hlinewd{1pt}

\rowcolor{BGTable} &  & \multicolumn{1}{c|}{\textbf{1.1}} & \multicolumn{1}{c|}{\textbf{1.2}} & \multicolumn{1}{c|}{\textbf{1.3}} & \textbf{1.4} & \multicolumn{1}{c|}{\textbf{2.1}} & \multicolumn{1}{c|}{\textbf{2.2}} & \textbf{2.3} & \multicolumn{1}{c|}{\textbf{3.1}} & \multicolumn{1}{c|}{\textbf{3.2}} & \textbf{3.3} \\ 
\hlinewd{1pt}

\cellcolor{BGTable} & \fullcite{ScharpfSG21b}; \cite{ScharpfSG21b} &  &  &  &  & \check & \check & \check &  &  &  \\ \hhline{>{{\vrule width 0.6pt}}|>{\arrayrulecolor{BGTable}}->{{\vrule width 0.6pt}\arrayrulecolor{black}}|-----------}
\multirow{-2}{*}[2em]{\cellcolor{BGTable}\textbf{2021}} & \fullcite{ScharpfSG21a}; \cite{ScharpfSG21a} &  &  &  &  & \check & \check &  &  &  &  \\ 
\hlinewd{1pt}

\rowcolor{BGTable} &  & \multicolumn{1}{c|}{\textbf{1.1}} & \multicolumn{1}{c|}{\textbf{1.2}} & \multicolumn{1}{c|}{\textbf{1.3}} & \textbf{1.4} & \multicolumn{1}{c|}{\textbf{2.1}} & \multicolumn{1}{c|}{\textbf{2.2}} & \textbf{2.3} & \multicolumn{1}{c|}{\textbf{3.1}} & \multicolumn{1}{c|}{\textbf{3.2}} & \textbf{3.3} \\ 
\hlinewd{1pt}

\cellcolor{BGTable} & \fullcite{GreinerPetterSMB20a}; \cite{GreinerPetterSMB20a} & \check &  & \check &  & \check &  &  &  &  &  \\ \hhline{>{{\vrule width 0.6pt}}|>{\arrayrulecolor{BGTable}}->{{\vrule width 0.6pt}\arrayrulecolor{black}}|-----------}
\cellcolor{BGTable} & \fullcite{AsakuraGAM20}; \cite{AsakuraGAM20} & \check &  & \check &  & \check &  & \check & \check &  & \check \\ \hhline{>{{\vrule width 0.6pt}}|>{\arrayrulecolor{BGTable}}->{{\vrule width 0.6pt}\arrayrulecolor{black}}|-----------}
\cellcolor{BGTable} & \fullcite{SchubotzGMT20a}; \cite{SchubotzGMT20a} &  & \check &  &  & \check & \check & \check &  & \check &  \\ \hhline{>{{\vrule width 0.6pt}}|>{\arrayrulecolor{BGTable}}->{{\vrule width 0.6pt}\arrayrulecolor{black}}|-----------}
\multirow{-4}{*}[5em]{\cellcolor{BGTable}\textbf{2020}} & \fullcite{GreinerPetterYRM20a}; \cite{GreinerPetterYRM20a} & \check &  &  &  & \check &  &  &  &  & \check \\ \hhline{>{{\vrule width 0.6pt}}|>{\arrayrulecolor{BGTable}}->{{\vrule width 0.6pt}\arrayrulecolor{black}}|-----------}
\hlinewd{1pt}
\end{tabularx}
\end{table}

\pagebreak
\begin{table}[H]
\centering
\setlength\tabcolsep{2.5pt}
\footnotesize
\begin{tabularx}{\textwidth}{?c?X?a|a|a|a?b|b|b?d|d|d?}
\hlinewd{1pt}

\rowcolor{BGTable} & \multicolumn{1}{c?}{} & \multicolumn{4}{c?}{\textbf{WP 1}} & \multicolumn{3}{c?}{\textbf{WP 2}} & \multicolumn{3}{c?}{\textbf{WP 3}} \\ \hhline{>{{\vrule width 0.6pt}}|>{\arrayrulecolor{BGTable}}->{{\vrule width 0.6pt}\arrayrulecolor{black}}|>{\arrayrulecolor{BGTable}}->{{\vrule width 0.6pt}\arrayrulecolor{black}}|----|---|---|} 
\rowcolor{BGTable}\multirow{-2}{*}{\textbf{Year}} & \multicolumn{1}{c?}{\multirow{-2}{*}{\textbf{Publication}}} & \multicolumn{1}{c|}{\textbf{1.1}} & \multicolumn{1}{c|}{\textbf{1.2}} & \multicolumn{1}{c|}{\textbf{1.3}} & \textbf{1.4} & \multicolumn{1}{c|}{\textbf{2.1}} & \multicolumn{1}{c|}{\textbf{2.2}} & \textbf{2.3} & \multicolumn{1}{c|}{\textbf{3.1}} & \multicolumn{1}{c|}{\textbf{3.2}} & \textbf{3.3} \\ 
\hlinewd{1pt}

\cellcolor{BGTable} & \fullcite{GreinerPetterSAG20}; \cite{GreinerPetterSAG20} & \check &  &  &  & \check & \check & \check & \check & \check & \check \\ \hhline{>{{\vrule width 0.6pt}}|>{\arrayrulecolor{BGTable}}->{{\vrule width 0.6pt}\arrayrulecolor{black}}|-----------}
\cellcolor{BGTable} & \fullcite{ScharpfSGO20}; \cite{ScharpfSGO20} &  &  &  &  & \check &  & \check &  &  &  \\ \hhline{>{{\vrule width 0.6pt}}|>{\arrayrulecolor{BGTable}}->{{\vrule width 0.6pt}\arrayrulecolor{black}}|-----------}
\multirow{-3}{*}[4em]{\cellcolor{BGTable}\textbf{2020}} & \fullcite{ScharpfSYH20a}; \cite{ScharpfSYH20a} &  &  &  &  & \check &  &  &  &  &  \\
\hlinewd{1pt}

\rowcolor{BGTable} &  & \multicolumn{1}{c|}{\textbf{1.1}} & \multicolumn{1}{c|}{\textbf{1.2}} & \multicolumn{1}{c|}{\textbf{1.3}} & \textbf{1.4} & \multicolumn{1}{c|}{\textbf{2.1}} & \multicolumn{1}{c|}{\textbf{2.2}} & \textbf{2.3} & \multicolumn{1}{c|}{\textbf{3.1}} & \multicolumn{1}{c|}{\textbf{3.2}} & \textbf{3.3} \\ 
\hlinewd{1pt}

\cellcolor{BGTable} & \fullcite{MeuschkeSSK19a}; \cite{MeuschkeSSK19a} & \check &  &  &  &  &  &  &  & \check & \check \\ \hhline{>{{\vrule width 0.6pt}}|>{\arrayrulecolor{BGTable}}->{{\vrule width 0.6pt}\arrayrulecolor{black}}|-----------}
\cellcolor{BGTable} & \fullcite{ScharpfMSB19a}; \cite{ScharpfMSB19a} &  &  &  &  & \check & \check & \check &  &  &  \\ \hhline{>{{\vrule width 0.6pt}}|>{\arrayrulecolor{BGTable}}->{{\vrule width 0.6pt}\arrayrulecolor{black}}|-----------}
\cellcolor{BGTable} & \fullcite{GreinerPetterRSA19a}; \cite{GreinerPetterRSA19a} & \check &  &  &  & \check &  &  &  &  & \check \\ \hhline{>{{\vrule width 0.6pt}}|>{\arrayrulecolor{BGTable}}->{{\vrule width 0.6pt}\arrayrulecolor{black}}|-----------}
\cellcolor{BGTable} & \fullcite{GreinerPetterSCG19a}; \cite{GreinerPetterSCG19a} &  &  & \check &  & \check & \check &  & \check &  & \check \\ \hhline{>{{\vrule width 0.6pt}}|>{\arrayrulecolor{BGTable}}->{{\vrule width 0.6pt}\arrayrulecolor{black}}|-----------}
\cellcolor{BGTable} & \fullcite{ScharpfSCG19a}; \cite{ScharpfSCG19a} &  &  &  &  &  & \check &  & \check &  &  \\ \hhline{>{{\vrule width 0.6pt}}|>{\arrayrulecolor{BGTable}}->{{\vrule width 0.6pt}\arrayrulecolor{black}}|-----------}
\multirow{-6}{*}[11em]{\cellcolor{BGTable}\textbf{2019}} & \fullcite{SchubotzTSM19a}; \cite{SchubotzTSM19a} &  &  &  & \check &  &  &  & \check &  & \check \\ 
\hlinewd{1pt}

\end{tabularx}
\end{table}

\pagebreak
\begin{table}[H]
\centering
\setlength\tabcolsep{2.5pt}
\footnotesize
\begin{tabularx}{\textwidth}{?c?X?a|a|a|a?b|b|b?d|d|d?}
\hlinewd{1pt}

\rowcolor{BGTable} & \multicolumn{1}{c?}{} & \multicolumn{4}{c?}{\textbf{WP 1}} & \multicolumn{3}{c?}{\textbf{WP 2}} & \multicolumn{3}{c?}{\textbf{WP 3}} \\ \hhline{>{{\vrule width 0.6pt}}|>{\arrayrulecolor{BGTable}}->{{\vrule width 0.6pt}\arrayrulecolor{black}}|>{\arrayrulecolor{BGTable}}->{{\vrule width 0.6pt}\arrayrulecolor{black}}|----|---|---|} 
\rowcolor{BGTable}\multirow{-2}{*}{\textbf{Year}} & \multicolumn{1}{c?}{\multirow{-2}{*}{\textbf{Publication}}} & \multicolumn{1}{c|}{\textbf{1.1}} & \multicolumn{1}{c|}{\textbf{1.2}} & \multicolumn{1}{c|}{\textbf{1.3}} & \textbf{1.4} & \multicolumn{1}{c|}{\textbf{2.1}} & \multicolumn{1}{c|}{\textbf{2.2}} & \textbf{2.3} & \multicolumn{1}{c|}{\textbf{3.1}} & \multicolumn{1}{c|}{\textbf{3.2}} & \textbf{3.3} \\ 
\hlinewd{1pt}

\cellcolor{BGTable} & \fullcite{CohlGS18a}; \cite{CohlGS18a} &  & \check & \check &  & \check & \check &  & \check &  & \check \\ \hhline{>{{\vrule width 0.6pt}}|>{\arrayrulecolor{BGTable}}->{{\vrule width 0.6pt}\arrayrulecolor{black}}|-----------}
\cellcolor{BGTable} & \fullcite{SchubotzGSM18a}; \cite{SchubotzGSM18a} &  & \check & \check & \check & \check & \check &  & \check &  & \check \\ \hhline{>{{\vrule width 0.6pt}}|>{\arrayrulecolor{BGTable}}->{{\vrule width 0.6pt}\arrayrulecolor{black}}|-----------}
\cellcolor{BGTable} & \fullcite{MeuschkeSSG18a}; \cite{MeuschkeSSG18a} &  &  &  &  &  &  &  &  & \check & \check \\ \hhline{>{{\vrule width 0.6pt}}|>{\arrayrulecolor{BGTable}}->{{\vrule width 0.6pt}\arrayrulecolor{black}}|-----------}
\cellcolor{BGTable} & \fullcite{GreinerPetterSCG18a}; \cite{GreinerPetterSCG18a} &  &  & \check &  &  &  &  & \check &  & \check \\ \hhline{>{{\vrule width 0.6pt}}|>{\arrayrulecolor{BGTable}}->{{\vrule width 0.6pt}\arrayrulecolor{black}}|-----------}
\cellcolor{BGTable} & \fullcite{PetersenSG18a}; \cite{PetersenSG18a} &  &  &  & \check &  &  &  & \check &  & \check \\ \hhline{>{{\vrule width 0.6pt}}|>{\arrayrulecolor{BGTable}}->{{\vrule width 0.6pt}\arrayrulecolor{black}}|-----------}
\cellcolor{BGTable} & \fullcite{ScharpfSG18a}; \cite{ScharpfSG18a} &  &  &  &  & \check & \check &  &  & \check &  \\ \hhline{>{{\vrule width 0.6pt}}|>{\arrayrulecolor{BGTable}}->{{\vrule width 0.6pt}\arrayrulecolor{black}}|-----------}
\cellcolor{BGTable} & \fullcite{Schubotz18a}; \cite{Schubotz18a} &  &  &  &  &  & \check &  &  &  &  \\ \hhline{>{{\vrule width 0.6pt}}|>{\arrayrulecolor{BGTable}}->{{\vrule width 0.6pt}\arrayrulecolor{black}}|-----------}
\cellcolor{BGTable} & \fullcite{Schubotz18GI}; \cite{Schubotz18GI} & \check & \check & \check & \check & \check & \check & \check &  & \check &  \\ \hhline{>{{\vrule width 0.6pt}}|>{\arrayrulecolor{BGTable}}->{{\vrule width 0.6pt}\arrayrulecolor{black}}|-----------}
\multirow{-9}{*}[14em]{\cellcolor{BGTable}\textbf{2018}} & \fullcite{SchubotzSDN18}; \cite{SchubotzSDN18} &  &  &  &  & \check & \check & \check &  &  &  \\ 
\hlinewd{1pt}

\rowcolor{BGTable} &  & \multicolumn{1}{c|}{\textbf{1.1}} & \multicolumn{1}{c|}{\textbf{1.2}} & \multicolumn{1}{c|}{\textbf{1.3}} & \textbf{1.4} & \multicolumn{1}{c|}{\textbf{2.1}} & \multicolumn{1}{c|}{\textbf{2.2}} & \textbf{2.3} & \multicolumn{1}{c|}{\textbf{3.1}} & \multicolumn{1}{c|}{\textbf{3.2}} & \textbf{3.3} \\ 
\hlinewd{1pt}

\cellcolor{BGTable}\textbf{2017} & \fullcite{MeuschkeSHS17}; \cite{MeuschkeSHS17} & \check &  &  &  &  &  &  &  & \check & \check \\ 
\hlinewd{1pt}
\end{tabularx}
\end{table}

\pagebreak
\subsection{Press Releases}\label{sec:press-releases}
The press relations office at the University of Wuppertal reported on the project's contributions to improving access to mathematical content in Wikipedia. The original press release is no longer available due to a complete overhaul and relaunch of the University's website. We attach a copy of the article (in German) below.
\vspace{-1cm}

\begin{figure}[H]
 \centering 
 \includegraphics[page=1,width=.95\textwidth]{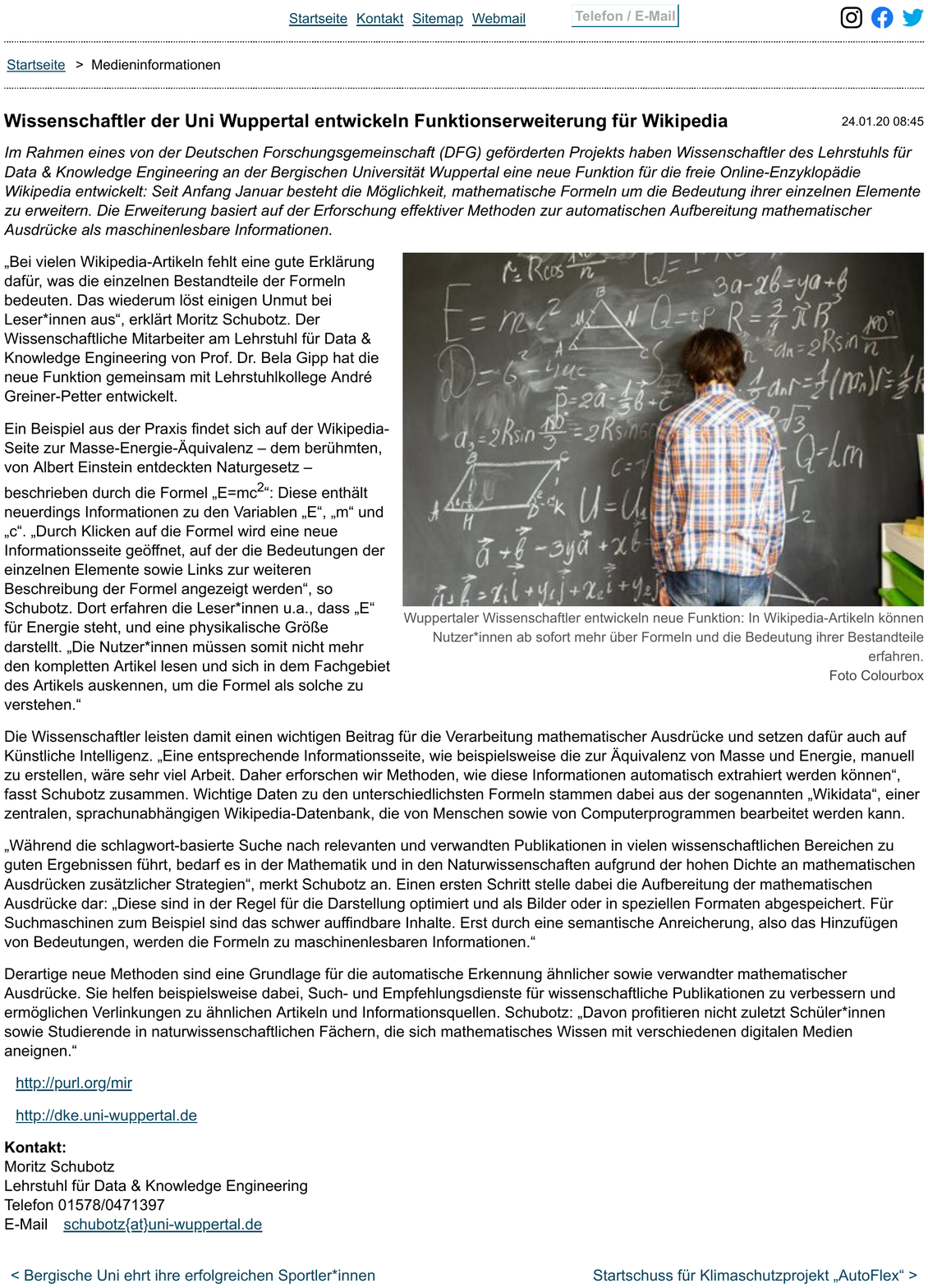}
 \vspace{-1cm}
\end{figure}


\section*{References}
\printbibliography[keyword=top,heading=none]\label{sec:pubs-top}

\vspace{-0.3cm}
{\color{PrimaryBlue}\rule{\textwidth}{0.5pt}}

\newrefcontext[sorting=nty]
\printbibliography[notkeyword=top,heading=none]\label{sec:pubs-rest}

\end{document}